\documentclass{sig-alternate-05-2015}

\newcommand{\fixnewfont}[6]{%
  \renewcommand#1{\fontsize{#5}{#6}\usefont{\encodingdefault}{#2}{#3}{#4}}%
}

\fixnewfont{\secfnt}{ptm}{b}{n}{12}{14}
\fixnewfont{\secit}{ptm}{b}{it}{12}{14}
\fixnewfont{\subsecfnt}{ptm}{m}{it}{11}{14}
\fixnewfont{\subsecit}{ptm}{b}{it}{11}{14}
\fixnewfont{\ttlfnt}{phv}{b}{n}{18}{20}
\fixnewfont{\ttlit}{phv}{b}{sl}{18}{20}
\fixnewfont{\subttlfnt}{phv}{m}{n}{14}{20}
\fixnewfont{\subttlit}{phv}{m}{sl}{14}{20}
\fixnewfont{\subttlbf}{phv}{b}{n}{14}{20}
\fixnewfont{\aufnt}{phv}{m}{n}{12}{16}
\fixnewfont{\auit}{phv}{m}{sl}{12}{16}
\fixnewfont{\affaddr}{phv}{m}{n}{10}{12}
\fixnewfont{\affaddrit}{phv}{m}{sl}{10}{12}
\fixnewfont{\eaddfnt}{phv}{m}{n}{12}{14}
\fixnewfont{\ixpt}{ptm}{m}{n}{9}{11}
\fixnewfont{\confname}{ptm}{m}{it}{8}{10}
\fixnewfont{\crnotice}{ptm}{m}{n}{8}{10}
\fixnewfont{\ninept}{ptm}{m}{n}{9}{10.5}

\usepackage{graphicx}
\usepackage{url}
\usepackage{amsmath}

\newcommand{\genasis}{{\sc GenASiS}}

\graphicspath{{figures/}}

\begin{document}

\title{Accelerating Our Understanding of Supernova Explosion Mechanism via Simulations and Visualizations with \textbf{\genasis} }

\numberofauthors{3}
\author{
\alignauthor
Reuben D. Budiardja\\
       \affaddr{National Institute for Computational Sciences}\\
       \affaddr{The University of Tennessee}\\
       \affaddr{Knoxville, Tennessee}\\
       \email{reubendb@utk.edu}
% 2nd. author
\alignauthor
Christian Y. Cardall\\
       \affaddr{Physics Division, Oak Ridge National Laboratory}\\
       \affaddr{Oak Ridge, Tennessee}\\ 
       \affaddr{Department of Physics and Astronomy, University of Tennessee}\\
       \affaddr{Knoxville, Tennessee}\\
       \email{cardallcy@ornl.gov}
% 3rd. author
\alignauthor Eirik Endeve\\
       \affaddr{Computer Science and Mathematics Division, Oak Ridge National Laboratory}\\
       \affaddr{Oak Ridge, Tennessee}\\ 
       \affaddr{Department of Physics and Astronomy, University of Tennessee}\\
       \affaddr{Knoxville, Tennessee}\\
       \email{endevee@ornl.gov}
}

\setcopyright{licensedusgovmixed}
\conferenceinfo{XSEDE '15,}{July 26 - 30, 2015, St. Louis, MO, USA}
\isbn{978-1-4503-3720-5/15/07}\acmPrice{\$15.00}
\doi{http://dx.doi.org/10.1145/2792745.2792746}

\maketitle

\begin{abstract}
Core-collapse supernovae are among the most powerful explosions in the Universe, releasing about $10^{53}~\mbox{erg}$ of energy on timescales of a few tens of seconds. These explosion events are also responsible for the production and dissemination of most of the heavy elements, making life as we know it possible. Yet exactly how they work is still unresolved. One reason for this is the sheer complexity and cost of a self-consistent, multi-physics, and multi-dimensional core-collapse supernova simulation, which is impractical, and often impossible, even on the largest supercomputers we have available today. To advance our understanding we instead must often use simplified models, teasing out the most important ingredients for successful explosions, while helping us to interpret results from higher fidelity multi-physics models. In this paper we investigate the role of instabilities in the core-collapse supernova environment. We present here simulation and visualization results produced by our code \genasis. 
\end{abstract}

% \begin{IEEEkeywords}
% Numerical Simulations; Hydrodynamics; Supernovae; 
% \end{IEEEkeywords}

\section{Introduction}
Core-collapse supernovae are the violent deaths of massive stars and are among the most powerful explosions in universe, releasing about $10^{53}~\mbox{erg}$ of energy on timescales of a few tens of seconds. This rivals the instantaneous power of all the rest of the luminous visible universe combined. They mark the birth of the most exotic states of matter known: neutron stars and black holes, while at the same time producing and disseminating most of the elements heavier than helium, making life as we know it possible. These events occur about twice per century in a typical galaxy like our own and have been in the forefront of research in the field of astronomy and astrophysics for almost half a century. Yet how exactly they work is still shrouded in mystery. 

Stars burn hydrogen into helium for most of their existence. For stars more massive than $\sim 10$ solar masses (${M}_\odot$), temperatures and densities are sufficiently high for burning to continue through carbon and to iron group elements. The star ends up in an onion-like configuration, with an iron core surrounded by layers of silicon, oxygen, carbon, helium, and hydrogen. Since the iron group elements are the most tightly bound, burning in the core ceases. At this point, the pressure in the core is dominated by electron degeneracy pressure (a consequence of the Pauli exclusion principle), which supports it against the inward pull of gravity. This balance between the gravitational pull and the electron degeneracy pressure in the core is only marginally stable. 

Two processes occur in the core that result in the reduction of the degeneracy pressure support: electron capture on the free protons and nuclei, and nuclear dissociation under extreme densities and temperatures. The pressure support in the core is reduced enough that the core eventually becomes unstable and collapses. As the core collapses, the inner and outer regions behave differently. The inner core undergoes homologous collapse---velocity increases linearly with radius---as expected of a fluid with relativistic, degenerate electron pressure. With increasing radius, the density decreases, and thus also the local sound speed. Thus, there is a radius where the speed of the infalling matter is the same as the local sound speed, demarcating the inner and outer core. Beyond this radius---the outer core---matter collapses supersonically. 

The inner core collapses until it exceeds nuclear matter density ($\sim 1 - 3 \times 10^{14}~\mbox{g}/\mbox{cm}^3$). At this extreme density, the pressure of the inner core increases dramatically as a result of the repulsive component of the short-range nuclear force. The inner core becomes incompressible and bounces, and a shock wave forms at the boundary of the inner and outer core and begins to move out. Ultimately this shock wave will be responsible for the disruption of the star, producing the observable explosion.

It was once thought that as the shock wave propagated outward, the velocity of the bounce would grow as it moved into the outer layers of the core; the bounce would therefore be the origin of the supernova's energy \cite{Colgate1960}. From all the more realistic models completed to date, we now know that this is not the case, and therein lies the core-collapse supernova problem. 

As the shock propagates out, it has to move through infalling material in the outer core, during which nuclear dissociation happens. This costs the shock energy. Additional energy losses occur when electron capture on the free protons liberated by nuclear dissociation eventually results in an electron neutrino burst.
As a result of these energy losses, the shock stalls. 

If this were the end of the story, no supernova would ever explode nor be observed. The shock has to be reenergized so that it may continue to propagate outward and eventually produce the explosion. The details of how the stalled shock is revived is the central question in core-collapse supernova theory. 

Out of the $10^{53}~\mbox{erg}$ released during the explosion, the visible explosion energy is only $1\%$. The rest is released as neutrinos. Because neutrinos dominate the energetics of a supernova event, it is natural to consider neutrino heating as a mechanism for the revival of the stalled shock. This delayed neutrino-heating has been proposed as one of the mechanisms that leads to explosion (for example, see \cite{Janka2012} for reviews). 

Core-collapse supernovae are asymmetric events. Observational evidence that has accumulated to support this includes spectropolarimetry, large average pulsar velocities, and the morphology of highly resolved images of supernova such as SN 1987A (see \cite{McCray1993} and references therein). On the theoretical side, simulations have shown that a variety of fluid instabilities are present. These instabilities develop convective overturn and help transport hot gas from neutrino-heating region directly to the shock, thereby enhancing the neutrino energy deposition to the stalled shock  (e.g \cite{Mezzacappa2005} and references). These multidimensional effects therefore may be important for the neutrino-heating mechanism to revive the stalled shock. Recent simulations have also revealed the existence of the standing accretion shock instability (SASI), which given enough time, may also grow and impact the dynamics \cite{Blondin2006}. All these multidimensional effects may play essential roles in possible mechanisms of core-collapse supernovae.

% Stars have both rotation and magnetic fields. It has been suggested that in more massive progenitors rotation and magnetic fields may play a more significant role \cite{Thompson2005,Fryer2000,Wheeler2006}, producing jet-like hypernovae, and perhaps giving birth to `magnetars', a type of neutron star with an unusually large magnetic field. Observational evidence seems to support this \cite{Gaensler2005,Figer2005}. Even in normal supernovae there seems to be observational evidence that rotation and magnetic field play some roles \cite{Burrows2004}. During collapse magnetic fields may also be amplified enough to have important dynamical effects. Recently we discovered that an amplification of magnetic fields can happen in the SASI (as discussed in chapter \ref{ch:sasi}), thereby extending the range of progenitors in which magnetic field may play a significant dynamical role.

Because of the complexity of the candidate supernova mechanisms, a purely analytic investigation is not possible. Instead, supernova modeling requires sophisticated numerical simulations. All the input physics required to model core-collapse supernovae present daunting challenges that are both algorithmic and computational in nature, and will tax state-of-the-art supercomputers for years to come. This is especially true for three-dimensional models and simulations, where multi-physics simulations with the required level of physical fidelity will not fit even the largest supercomputers currently available. Instead, in three spatial dimensions we must often resolve to simplified models to tease out the important ingredients for successful supernova explosions. There is indeed a value to this approach as it also helps us to interpret results from the much more computationally expensive multi-physics models.

The investigation reported in this paper follows the same approach. In particular, we are interested to learn more about the role of hydrodynamics instabilities in core-collapse supernova environments. We build on previous results by \cite{Fernandez2014} and extend their work to three spatial dimensions. This paper is organized as follows. We first describe our numerical code in section \ref{sec:intro}. In section \ref{sec:investigation} we describe our simulations and results, followed by summary and conclusion in section \ref{sec:conclusion}. All of our simulations were done using the Darter supercomputer \cite{Fahey2014} at the National Institute for Computational Sciences (NICS). 

\section{The Tool: \genasis\ }
\label{sec:intro}
\genasis\  (\textit{Gen}eral \textit{A}strophysical \textit{Si}mulation \textit{S}ystem) is a new code we are developing to facilitate simulations of astrophysical problems on high-performance computing resources. The word `General' in its name denotes its extensibility to implement and use multiple physics, solvers, and numerical algorithms by virtue of abstracted names and/or interfaces. Rather than a single program, it is a `Simulation System' comprising a collection of modules, structured as classes, which suitable driver programs can use to initialize and solve particular problems of interest. `Astrophysical' denotes the kinds of applications the code is ultimately aimed at and the types of solvers it will eventually provide. In this section we briefly discuss some features of \genasis\ relevant to the simulations at hand. 
%We begin by some discussion of \genasis\ code design.

% \subsection{Code Design}
% Many problems in astrophysics and cosmology are multi-physics in nature, where inputs from different physical phenomena are in play, commonly through a system of of partial differential equations. 
% %Proper modeling of these events requires a code with multi-physics input. 
% One prime example of such problems includes the simulations of core-collapse supernovae whose proper modeling requires solvers for hydrodynamics, magnetic fields, self-gravity, relativity, radiation transport, and nuclear physics. Building simulation codes capable to simulate all the desired physics right of the bat is likely not feasible. A more prudent approach to take is to deploy over time a series of approximations of increasing sophistication deployed. One primary motivator for this approach is the fact that for such complex phenomena, the level of physical fidelity  we can simulate is often constrained by the computational power available. As an example, 
% 
%  the multi-physics 

\subsection{Numerical Hydrodynamics}

The equations of ideal Newtonian hydrodynamics in conservative Eulerian formulation take the form

\begin{eqnarray}
\frac{\partial D}{\partial t} 
  + \frac{\partial}{\partial x^{i}} \left(\rho v^i \right) & = & 0,
\label{eq:HD_Mass} \\
\frac{\partial S^j}{\partial t} 
  + \frac{\partial}{\partial x^{i}} \left(\rho v^{j}v^{i} + p \delta^{ij} \right) & = & \mathcal{S}_M^j,
\label{eq:HD_Momentum} \\
\frac{\partial E}{\partial t} 
  + \frac{\partial}{\partial x^{i}} 
     \left(\left[e + p +\frac{1}{2}\rho v^j v^j\right]v^{i} \right)
  & = & \mathcal{S}_E.
\label{eq:HD_Energy}
\end{eqnarray}

These are the equations of mass, momentum, and energy conservation, respectively. In these equations, $\rho$, $v^i$, $e$, and $p$ represent rest mass density, fluid velocity, internal energy density, and fluid pressure, while $\mathcal{S}_M^j$ and $\mathcal{S}_E$ represent momentum and energy source terms. 
%Here we have also used Einstein's summation convention to denote spatial three-dimensional vectors, e.g. to collapse three equations for the three spatial dimensional to one in equation (\ref{eq:HD_Momentum}). 
Summation over repeated indices is implied.
The time-evolved \textit{conserved} quantities $D$, $E$, and $S^j$ are related to the \textit{comoving} variables $\rho$, $e$, $v^i$ through the relations
\begin{eqnarray}
D &=& \rho, \label{eq:HD_conserved_mass} \\
S^j &=& \rho v^j, \label{eq:HD_conserved_momentum} \\
E &=& e + \frac{1}{2} \rho v^j v^j. \label{eq:HD_conserved_energy}
\end{eqnarray}

The system of equations is closed by an equation of state relating the pressure to the internal energy. For an ideal gas, a polytropic equation of state of the form 
\begin{equation}
p = e (\Gamma - 1)
\label{eq:ideal_gas_eos}
\end{equation}
is used where $\Gamma$ is the adiabatic index.

In \genasis\ we have implemented a solver for Equations (\ref{eq:HD_Mass}) - (\ref{eq:HD_Energy}) based on the finite volume method. This is done by casting the equations as a system hyperbolic conservation laws of the form
\begin{equation}
\frac{\partial \mathbf{u}}{\partial t} + \nabla \cdot \mathbf{f} = 0,
\label{eq:conservation_law}
\end{equation}
where the vector of the conserved variables $\mathbf{u}$ and their corresponding fluxes $\mathbf{f}^i$ are defined as
\begin{eqnarray}
\mathbf{u} &=& \left[D, S^j, E\right], \label{eq:HD_conserved_vars}  \\
\mathbf{f}^i &=& \left[\rho v^i, \rho v^j v^i + p\delta^{ij}, 
                       \left(e + p +\frac{1}{2}\rho v^j v^j\right)v^{i} \right].
\label{eq:HD_raw_fluxes}
\end{eqnarray}

The fluxes through cell faces are computed using the so-called HLL-type approximate Riemann solvers \cite{Harten1983}\cite{Toro1994}. We have implemented two variants of these solvers. The first one, which we simply label `HLL'---for Harten, Lax, van Leer who first devised the method \cite{Harten1983}---is given by the expression

\begin{equation}
\mathbf{f}^{\mathrm{HLL}} 
  = \frac{\alpha^+ \mathbf{f}^\mathrm{L} + \alpha^- \mathbf{f}^\mathrm{R} - 
            \alpha^+\alpha^-\left(\mathbf{u}^\mathrm{R}-\mathbf{u}^\mathrm{L}\right)}
         {\alpha^+ + \alpha^-}. 
\label{eq:fluid_HLL_flux}
\end{equation}
The superscripts $\mathrm{L}$ and $\mathrm{R}$ denote the values on the left and right sides of the cell face, respectively, as given by equations (\ref{eq:HD_conserved_vars}) - (\ref{eq:HD_raw_fluxes}). The coefficients $\alpha^{\pm}$ are constructed from the maximum and minimum characteristic speeds $\lambda^{\pm}$ as \begin{equation}
\alpha^{\pm} = \max\left\{0, \pm \lambda^{\pm}\left(\mathbf{v}^\mathrm{L}\right),
                           \pm \lambda^{\pm}\left(\mathbf{v}^\mathrm{R}\right)
                    \right\}, 
\end{equation}
where
\begin{equation}
\lambda^{\pm} = v \pm \sqrt{\Gamma p / \rho}.
\end{equation}

\begin{figure}
\centering
\includegraphics[width=0.23\textwidth]{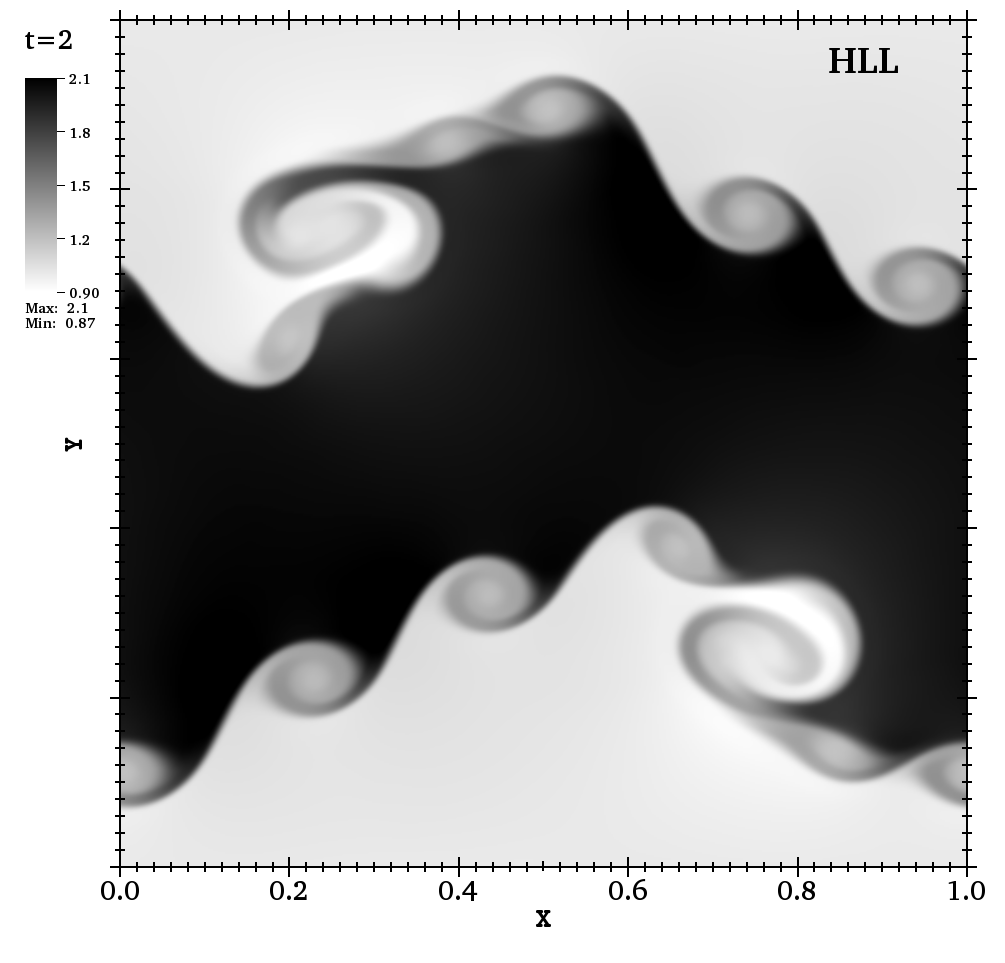}
\includegraphics[width=0.23\textwidth]{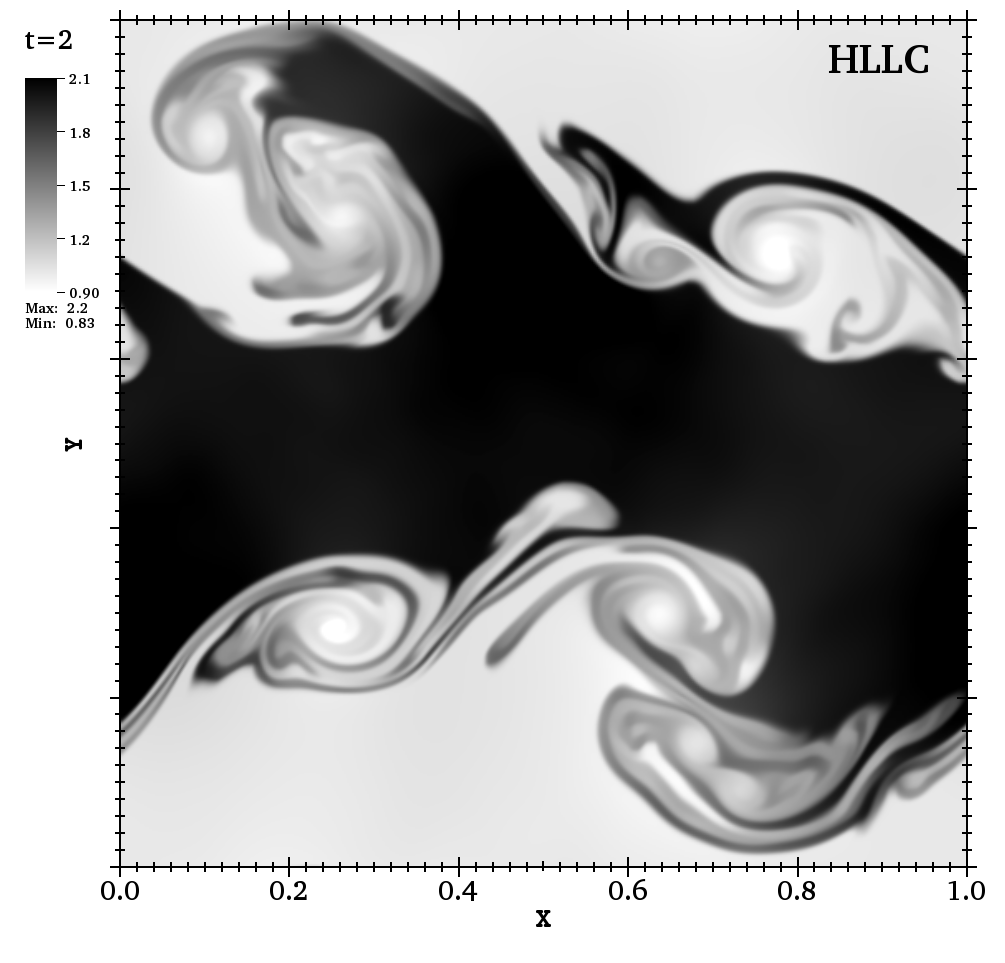}
\caption{Density plots of Kelvin-Helmholtz instability test as computed using HLL (left) and HLLC (right) solver on $512^2$ grid. The HLLC solver produces a more accurate and complex structure of the instability.}
\label{fig:kelvin_helmholtz}
\end{figure}

The second HLL Riemann solver variant we have implemented, known as the HLLC solver \cite{Toro1994}, takes into account both the acoustic and entropy waves which results in higher accuracy in preserving contact discontinuities. 
%FIXME: The HLLC flux is given by
Our implementation of the HLLC solver includes a fallback to the more dissipative HLL solver in grid-aligned shocks to avoid ``odd-even decoupling'' \cite{Quirk1994}. We have verified our implementation of these solvers using various numerical test problems with known solutions. These are detailed in \cite{Cardall2013}. %As an illustration of the differences between the HLL and HLLC solvers, we show in 
Figure \ref{fig:kelvin_helmholtz} shows the effect of the HLLC solver relative to the HLL solver on the Kevin-Helmholtz instability test problem.

In \genasis\ the conserved and comoving variables are defined as cell-centered values and must be reconstructed on the cell faces for the flux computation. We use slope-limited linear interpolation (which for cells of equal size yields a spatially second-order accurate scheme for smooth flows) to preserve non-oscillatory behavior near shocks and discontinuities. Specifically we have implemented a one-parameter ($\theta\in[1,2]$) family of generalized minmod limiters \cite{Kurganov2000}, which for an arbitrary variable $\chi$ is given by 
\begin{equation}
\left.\frac{\partial \chi}{\partial x}\right|_{ijk}
 = \mbox{minmod}~
   \left[
   \begin{array}{ll}
   \theta\left(\frac{\chi_{ijk} - \chi_{i-1jk}}{x_i - x_{i-1}}\right),\\
   \left(\frac{\chi_{i+1jk} - \chi_{i-1jk}}{x_{i+1} - x_{i-1}}\right),\\
   \theta\left(\frac{\chi_{i+1jk} - \chi_{ijk}}{x_{i+1} - x_{i}}\right)
   \end{array}
   \right] 
%   \theta\left(\frac{\chi_{ijk} - \chi_{i-1jk}}{x_i - x_{i-1}}\right),\\
%   \left(\frac{\chi_{i+1jk} - \chi_{i-1jk}}{x_{i+1} - x_{i-1}}\right),\\
%   \theta\left(\frac{\chi_{i+1jk} - \chi_{ijk}}{x_{i+1} - x_{i}}\right)
%   \right],
\label{eq:generalized_minmod}
\end{equation}
where the multi-variable minmod function is defined as
\begin{equation}
\mbox{minmod}\left(\chi_1, \chi_2, \dots\right) 
  = \left\{ 
      \begin{array}{ll}
      \mbox{min}_j\left(\chi_j\right),\qquad \mbox{if }~ \chi_j > 0 ~\forall j, \\
      \mbox{max}_j\left(\chi_j\right),\qquad \mbox{if }~ \chi_j < 0 ~\forall j, \\
      0                   \qquad\qquad\qquad \mbox{otherwise}. 
      \end{array}
    \right.
\end{equation}
In the simulations presented here we have used $\theta=1.4$. 

For time-stepping, we use a second-order TVD Runge-Kutta scheme \cite{Shu1988}:
\begin{eqnarray}
u^{(1)} &=& u^n + \Delta t L(u^n) \\
u^{(n+1)} &=& \frac{1}{2}u^n + \frac{1}{2}u^{(1)} + \frac{1}{2} \Delta t L(u^{(1)}),
\end{eqnarray}
where the operator $L(u^n)$ denotes the spatial differencing. The time interval $\Delta t$ must obey the Courant-€"Friedrichs€"-Lewy (CFL) condition given by 
\begin{equation}
v_c \cdot \Delta t < C \Delta x,
\end{equation}
where $v_c$ is the maximum characteristic speed on the grid, $\Delta x$ is the cell spacing, and $C < 1$ is the ``Courant parameter.'' 

\subsection{Multilevel Mesh Refinement}

Adaptive mesh refinement (AMR) is a technique to dynamically furnish higher resolution only when and where the problem requires it, thereby enabling focused deployment of limited computational resources. This technique becomes increasingly important in multi-physics codes since advances in scientific modeling, algorithms, and physical fidelity have so far always exceeded resources that can be provided even by the largest supercomputers. 

There are two common approaches to AMR. Block-structured AMR \cite{Berger1989}, as the name implies, uses blocks of cells as the basic unit of the mesh. The coarsest grid, consisting of several blocks, covers the entire computational domain. Each block may be refined, creating finer nested grids. 

In developing \genasis\ however we have instead chosen to use cell-by-cell AMR \cite{Khokhlov1998}. In this approach, individual cells may be refined (and coarsened) as needed; this fine-grained control leads to a smaller number of total cells and therefore potentially larger savings of computational resources. Our motivation for this approach is based on the very high computational cost for each spatial cell in the types of models we eventually intend to run. For example, in its full glory the solver for the neutrino radiation transport equations will have to compute the coupled interactions in momentum space for every spatial cell. In this case, the flexibility of cell-by-cell AMR may prove to be advantageous overall.

\begin{figure}
\centering
\includegraphics[width=0.235\textwidth]{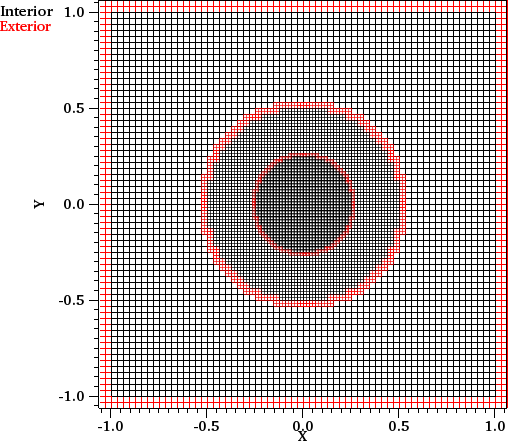}
\includegraphics[width=0.235\textwidth]{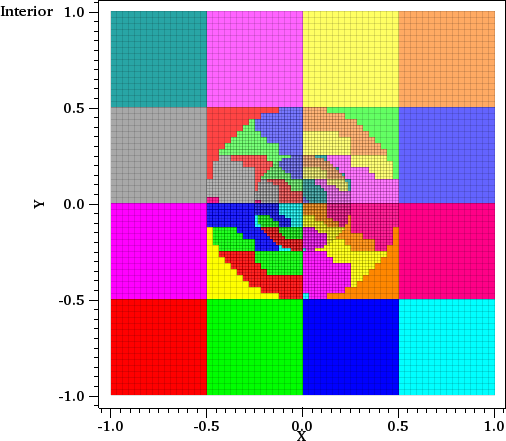}
\caption{Illustrations of a multilevel mesh with refinement for a total of three levels (left) and their domain decomposition (right) where different processes are indicated by different colors. The red cells (left) show the exteriors of the level meshes, which is the computational boundary at the coarsest level and the coarse/fine boundary on the finer levels. On the finer levels, the exterior values are prolongated from the coarser levels.}
\label{fig:level_domain_decomposed}
\end{figure}

In our AMR implementation, we take a level-by-level approach in arranging storage and solvers. Each level of refinement is a mesh in its own right. The meshes at different levels of refinement are therefore domain-decomposed independently for parallel processing (see Figure \ref{fig:level_domain_decomposed} for illustration). This approach simplifies the formulation of solvers since they only need to be written for a single level, while interactions between levels are handled separately. In our level-by-level approach, solutions from finer levels are used to improve solutions on coarser levels via restriction operations, and solutions from coarser levels are used as boundary conditions on finer levels via prolongation operations. Further details of our level-by-level formulation are given in \cite{Cardall2013}.

% \begin{figure}
% \centering
% \includegraphics[width=0.235\textwidth]{Sedov_2D_FMR_01}
% \includegraphics[width=0.235\textwidth]{Sedov_2D_FMR_04}
% \caption{}
% \label{fig:sedov_2d}
% \end{figure}

\subsection{Parallelization}

\genasis\ uses a simple `brick' decomposition to domain-decompose the computational domain into several subdomains. The brick decomposition simply means that, in three dimensions, the computational domain is divided in each dimension by $n_b = \sqrt[3]{n_p}$, i.e. the cube root of the number of processors $n_p$. Each subdomain is then assigned to an MPI process.  

For solvers requiring nearest-neighbor cell values crossing process boundaries, each subdomain keeps `ghost' cells in addition to the `proper' cells (the normal working computational cells assigned to the process). The values in ghost cells are then populated via point-to-point message passing with neighboring processes that own the corresponding proper cells. The brick decomposition maximizes the ratio of proper-to-ghost cells, therefore minimizing the nearest-neighbor communications.

\begin{figure}
\centering
\includegraphics[width=0.4\textwidth]{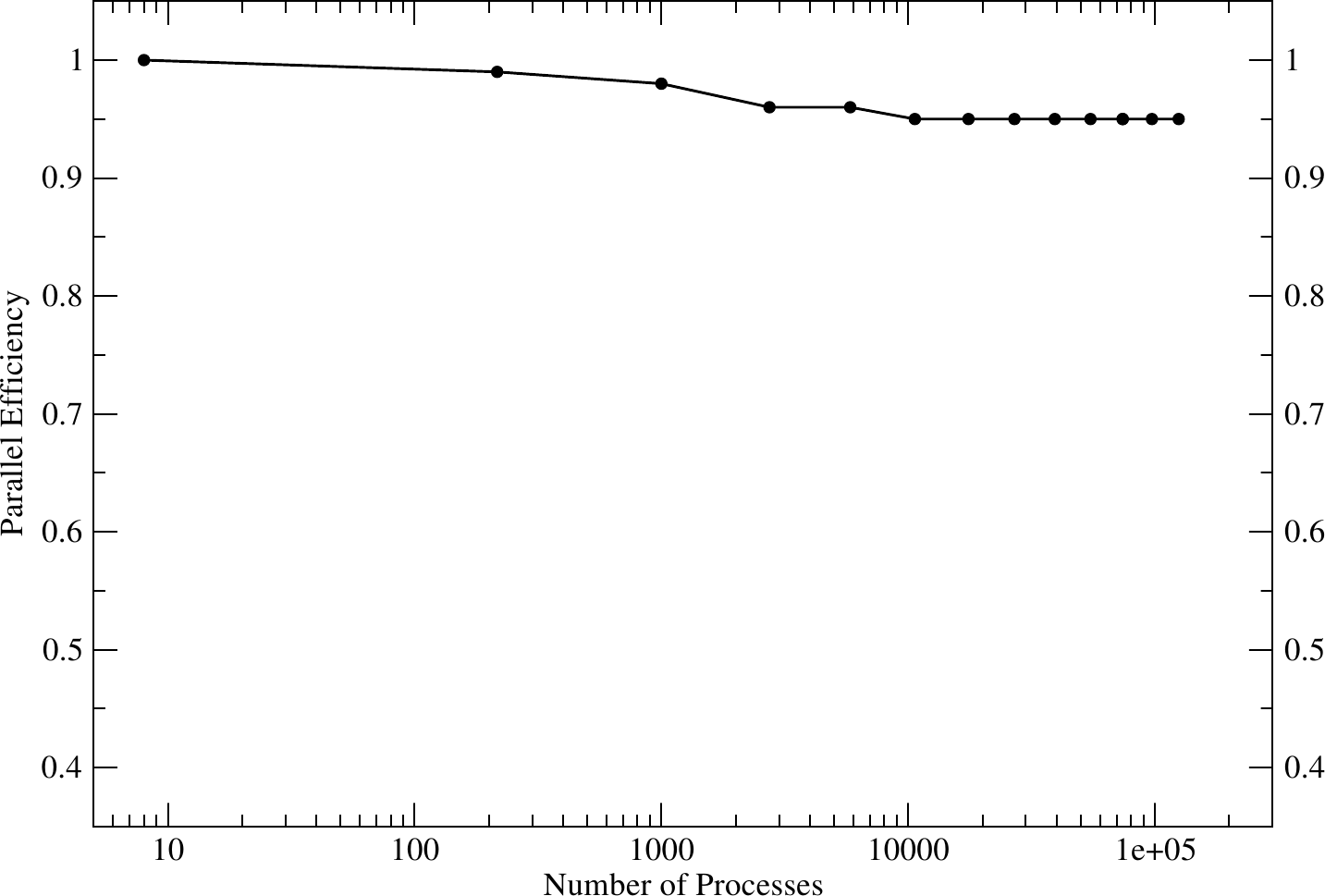}
\caption{MPI weak scaling of the hydrodynamics solver in GenASiS with a single-level mesh, where the number of computational cells per MPI process is kept fixed to $48^3$ as the number of processes are increased. High efficiency of around $0.95$ is maintained up to $100,000$ MPI processes.}
\label{fig:weak_scaling}
\end{figure}

To facilitate these exchanges of ghost cell data, we created two sets of ``connectivity'' for the proper cells. The ``exchange'' connectivity contains proper cells whose values need to be communicated to neighboring processes to populate their ghost cells, while the ``non-exchange'' connectivity contains the rest. Required computations are performed first on the exchange cells, after which a non-blocking, point-to-point MPI communication is initiated. While communication continues in the background, computations are performed on the non-exchange cells. This technique of overlapping work and communication proves to be essential to achieve the high-scalability of our code, as shown in Figure \ref{fig:weak_scaling}.

\section{The Investigation: The Role of Convection and SASI}
\label{sec:investigation}

\subsection{Initial Conditions}
We adopt a simplified model of the post-bounce core-collapse supernova environment in three dimensions. We use an ideal gas equation of state (c.f. equation \ref{eq:ideal_gas_eos}) with $\Gamma = 4/3$. Our model accounts for (momentum and energy) sources due to central gravity, heating and cooling due to neutrino radiation ($Q_v$), and nuclear dissociation/recombination ($Q_D$) as  
\begin{eqnarray}
\mathcal{S}_M^j & = & -\rho \frac{\partial \Phi}{\partial x^{j}} \\
\mathcal{S}_E & = & -\rho v^i \frac{\partial \Phi}{\partial x^{i}} + Q_v + Q_D, \label{eq:energy_source}
\end{eqnarray}
where $\mathcal{S}_M^j$ and $\mathcal{S}_E$ enter into the hydrodynamics equations as given by Equations (\ref{eq:HD_Momentum})-(\ref{eq:HD_Energy}). We use the point mass gravitational potential $\Phi = - GM/r$ where $G$ is the gravitational constant, $M$ is the mass of the central compact object (i.e. the proto-neutron star), and $r$ is the radial distance from the center of the star. Throughout our simulations we set $M = 1.3\ M_\odot$, where $M_\odot$ is the Solar mass. 

We follow \cite{Fernandez2014} in parameterizing the heating and cooling effects due to neutrino radiation, using values for the parameters that roughly mimic those in self-consistent supernova simulations. The heating and cooling function is given by
\begin{equation}
Q_v = \rho \left(\frac{B}{r^2} - A p^{3/2}\right) e^{-(s/s_0)^2} H(\mathcal{M}_0-\mathcal{M}),
\end{equation}
where $B$ is a parameter proportional to the fixed neutrino luminosity (i.e. heating) as derived in \cite{Janka2000}. The values for the cooling parameter $A$ are determined empirically as those which yield zero fluid velocity at the radius of the proto-neutron star $R_{\rm{PNS}}$ for a shock at radius $R_{\rm{sh}}=100$ km with no heating (i.e $B=0$). We set $R_{\rm{PNS}}=40$ km in all of our simulations as the fixed inner boundary.    $s\propto\log(p/\rho^\Gamma)$ is the entropy with the referential value $s_0$ for the minimum initial post-shock entropy. The Heaviside step function $H(\mathcal{M}_0-\mathcal{M})$ excludes heating and cooling outside of the shock based on the Mach number $\mathcal{M}$ with $\mathcal{M}_0=2$.

We include a nuclear dissociation energy per unit mass $\epsilon$ in the shock jump conditions which in turn influences whether the instability is \textit{SASI-dominated} or \textit{convection-dominated}. Larger values of $\epsilon$ corresponds to larger shock compression ratios and slower post-shock speeds. %FIXME(CITATION)
For these simulations, in addition to equations (\ref{eq:HD_Mass})-(\ref{eq:HD_Energy}) we also solve for the mass density of the dissociated baryons $\rho_D$:
\begin{equation}
\frac{\partial \rho_D}{\partial t} + \mathbf{\nabla} \cdot (\rho_D \mathbf{v}) = R_D,
\end{equation}
where the source term $R_D$ is given by
\begin{eqnarray} 
R_D &=& \frac{\rho - \rho_D}{\Delta t}\ \ \ \ (\mathrm{dissociation}), \\
&=&  -\frac{\rho_D}{\Delta t}\ \ \ \ \ \ (\mathrm{recombination}).
\end{eqnarray}
For numerical stability we spread the change over several time steps by taking $R_D \rightarrow f R_D$, where $f = \mathrm{min} ( 0.1, f_e )$, in which $f_e$ is the value that yields a 1\% change in internal energy density due to dissociation. The associated energy source term due to dissociation is then 
\begin{equation}
Q_D =  -\epsilon R_D.
\end{equation}

\subsection{Simulation Setup}
\label{subsec:simulations}
We produced two series of models: a SASI-dominated series with $\epsilon=0$ (no dissociation), and a convection-dominated series with $\epsilon=0.3$. For each series we increased the heating value $B$ until explosions were attained. For each value of $B$ the initial shock radius $R_{\rm{sh}}$ was determined as that which yielded zero fluid velocity at $R_{\rm{PNS}}$ using the previously obtained (with zero heating) value of $A$. We used Mathematica\footnote{http://www.wolfram.com/mathematica/ version 10.0} to obtain initial profiles, solving spherically symmetric steady-state versions of equations (\ref{eq:HD_Mass})-(\ref{eq:HD_Energy}) with the Rankine-Hugoniot shock jump conditions joining the pre- and post-shock regions. 

The initial profile was then mapped to our three-dimensional multi-level mesh with refinement, covering the computational domain $\left[-640, 640 \right]$ km. A total of four mesh levels was used. The coarsest level mesh had a resolution of $128^3$ covering the entire computational domain, yielding a spatial resolution of $10$ km. Each subsequent level increased the resolution by a factor of two in each dimension. The finest level of mesh therefore had a resolution of $1.25$ km. $64$ MPI processes were used to run each model in parallel. The computational cost for a single model was about $12,000$ SU on Darter.

%FIXME: . Level 2 mesh covers the domain $[-320, 320]$; level 3 mesh covers the domain  
%FIXME:(NEED GRID INFORMATION \& RESOLUTION, PROCS)

The initial condition was perturbed by introducing random pressure perturbations of amplitude $0.1$\% relative to the local value. The perturbations seeded the instabilities. For each value of heating for each series, we ran ten simulations of the same model (differing only by their initial random perturbations). We let each simulation run to either $1$ second of physical time or until explosion was attained. 

Most of the simulations presented here were run in the last few weeks of the year 2014, taking advantage of Darter's lower utilization rate due to the holiday. To get better job throughput for our simulations, and for easier job management, we bundled the ten simulations in each model into a single job request on Darter. Furthermore, when computational resources were available, we spread out the rank placement of our jobs, using only four to six MPI processes per socket. This provided higher memory bandwidth and effectively boosted the computational speed by 30 to 40 percent, since the hydrodynamics algorithm is inherently memory-bandwidth bound.

\subsection{Results}
\begin{figure}
\centering
\includegraphics[width=0.46\textwidth]{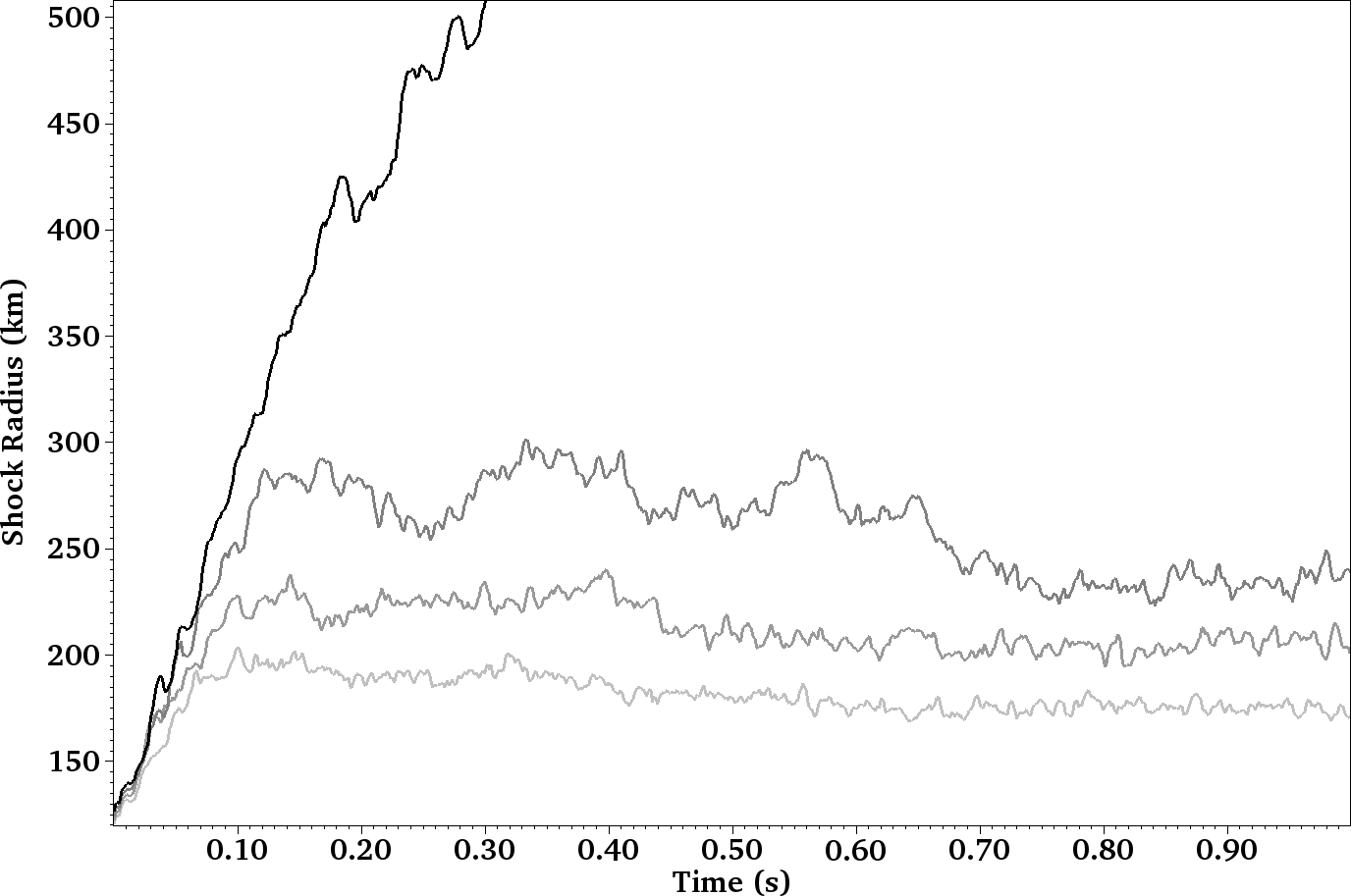}
\caption{Shock trajectories of representative convection-dominated models with heating values $B=[0.65,~0.70,~0.75,~0.80]$, where darker color indicates higher $B$. Only one of the ten simulations for each value $B$ is plotted.}
\label{fig:convection_dominated_models}
\end{figure}

\begin{figure}
\centering
\includegraphics[width=0.46\textwidth]{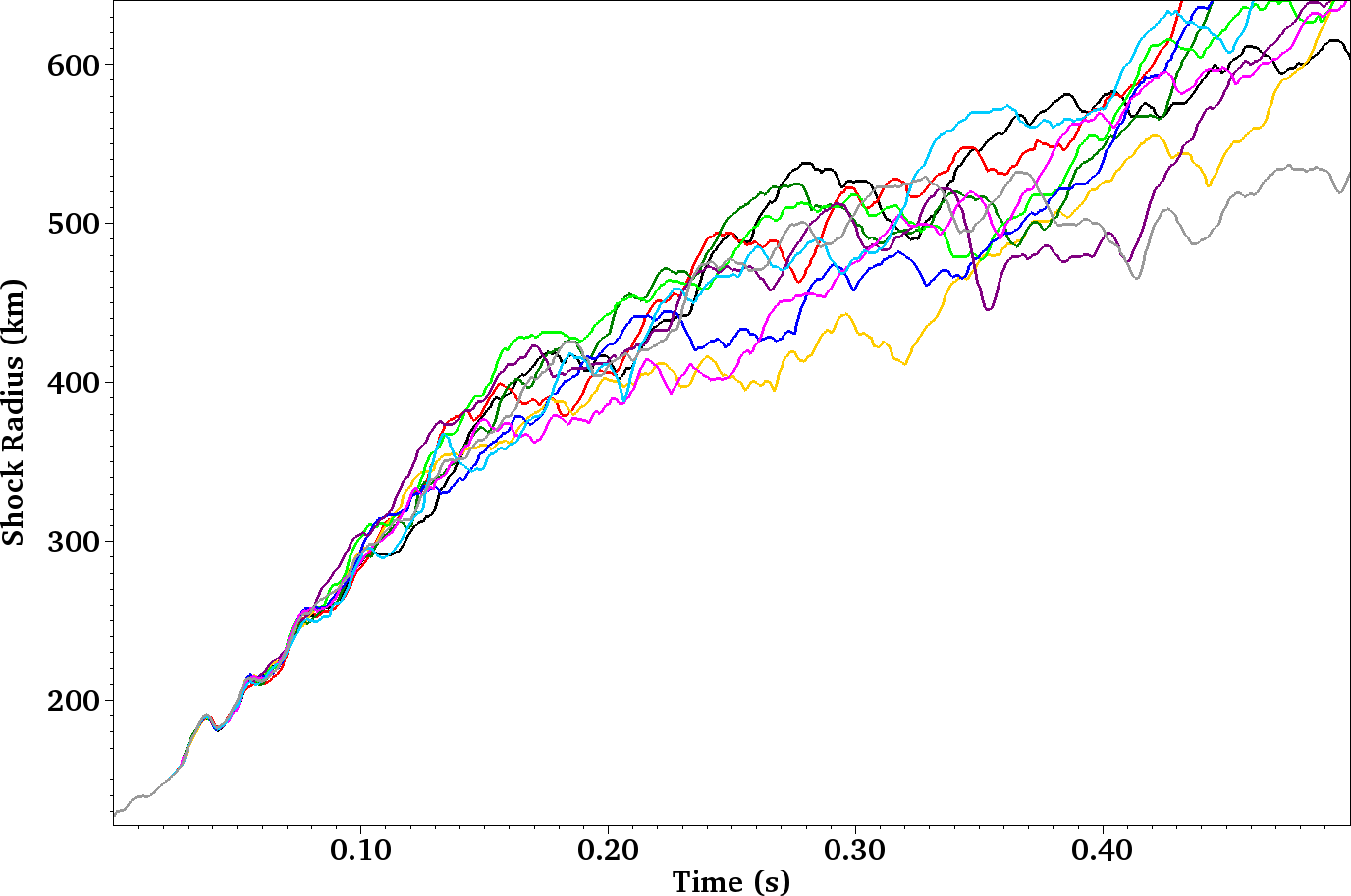}
\caption{Shock trajectories of ten simulations (indicated by different colors) at heating value $B=0.80$ for the convection-dominated model.}
\label{fig:convection_dominated_explosions}
\end{figure}

For the series of convection-dominated models $(\epsilon=0.3)$, as we increase the heating values $B$ the shock expands and settles at larger radius. Figure \ref{fig:convection_dominated_models} shows the shock radii of some representative convection-dominated models using various heating values $B$. 
For these models, there seems to be a critical heating value $B$ for which explosions are attained. In our simulations this value is between $B=0.75$ and $B=0.80$. All the convection-dominated models at and below $B=0.75$ did not attain explosion, while all the models at and above $B=0.80$ attained explosions. Figure \ref{fig:convection_dominated_explosions} shows the shock radii of the convection-dominated models for $B=0.80$. Although the shock radii do not follow exactly the same trajectories (due to different perturbations of the initial conditions), all of the simulations for this model have roughly similar explosion time scales.

\begin{figure}
\centering
\includegraphics[width=0.46\textwidth]{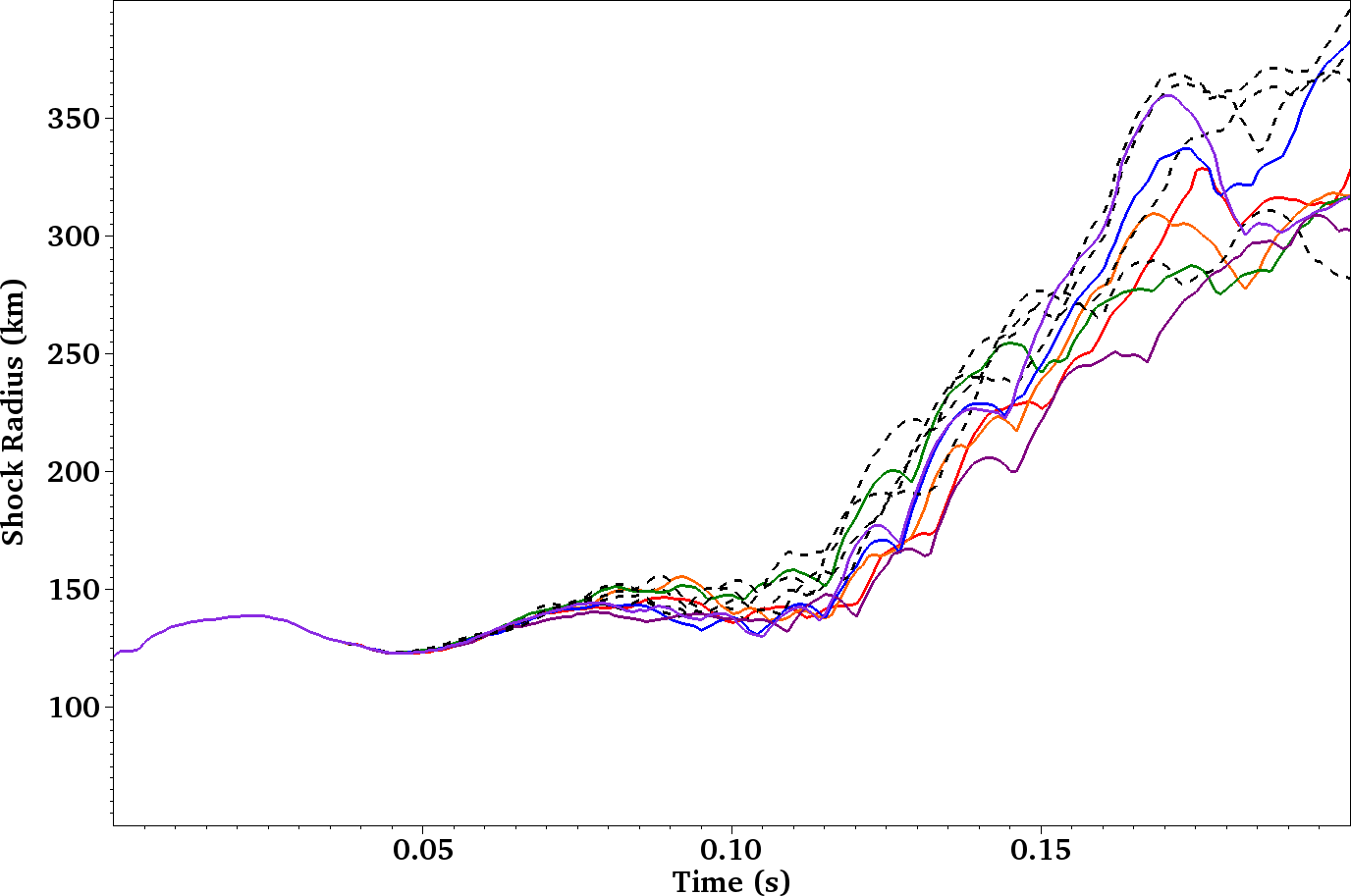}
\includegraphics[width=0.46\textwidth]{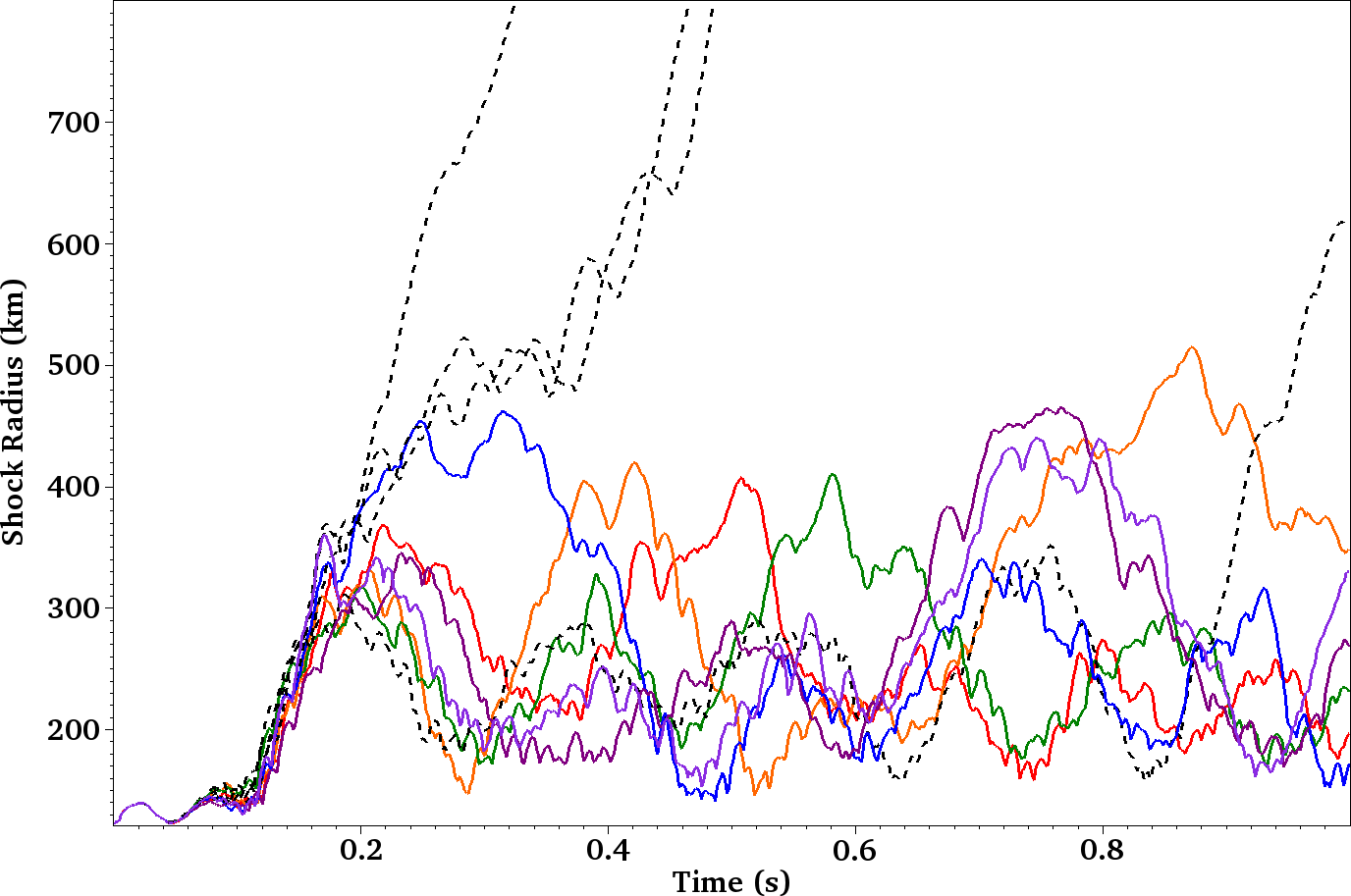}
\caption{Shock trajectories of ten simulations for model at heating value $B=1.025$ for SASI-dominated series. The upper panel shows a zoom-in of the trajectories at early time.}
\label{fig:SASI_dominated_models}
\end{figure}

The series of SASI-dominated models tells a completely different story. For these models, there does not seem to be a critical heating value $B$ for which explosions are guaranteed. Rather, as we increase the heating values $B$, \emph{only the probability of explosion increases}. Figure \ref{fig:SASI_dominated_models} plots the shock radii of ten SASI-dominated simulations for the model with heating value $B=1.025$. Exploding simulations are plotted with dash-lines, while non-exploding ones are plotted with solid lines of different colors to indicate the different simulations. In the upper panel of Figure \ref{fig:SASI_dominated_models}, a \textit{zoomed-in} version of the plot is shown for the first $200$ milliseconds of the simulations. The only differences in these simulations are the small random perturbations in their initial conditions. As seen on the upper panel of Figure \ref{fig:SASI_dominated_models}, for at least the first $50$ milliseconds, all simulations follow a very similar path in their shock radius plots. By $100$ milliseconds, the simulations have already begun diverging in their evolution. In the lower panel of Figure \ref{fig:SASI_dominated_models}, we see that eventually four simulations attained explosions for this heating value while the rest did not. Furthermore, for the exploding simulations, explosions were reached at vastly different time scales: one exploding simulation reached critical shock radius of $600$ kilometers at around $250$ milliseconds, two exploding simulations reached it at around $500$ milliseconds while maintaining much of the oscillatory behavior characteristic of SASI, and one simulation reached it at a much later time of nearly one second of evolution after a long period of vigorous SASI activity.

%Figure \ref{SASI_histogram} shows the histogram of 
   
\subsection{Visualizing the Data}

Scientific visualization is an integral part of many scientific endeavors, helping us glean valuable first insights from data generated by the simulations. The higher dimensionality nature of the supernova problem in particular---three dimensions in position space plus another three dimensions in momentum space in radiation transport---and the size of the data from each simulation often push the innovation and capability of scientific visualization. For the simulations discussed here however, although each simulation produces only a modest amount of data (order of $2$ Terabytes per simulation), the number of simulations done and the number of images and movies generated require us to develop and automate some kind of visualization workflow, which we discuss here.

\genasis\ has interfaces to the Silo library\footnote{https://wci.llnl.gov/simulation/computer-codes/silo} to be used for both checkpoint-restart files and output files. Silo was chosen  for its portability and because it is readily readable by the visualization package VisIt,\footnote{https://wci.llnl.gov/simulation/computer-codes/visit/} which we have been using extensively.

We use entropy to visualize the fluid, with higher entropy indicating hotter fluid. To help  visualize the three-dimensionality of the problem, two-dimensional slices parallel to the $xy-$, $xz-$, and $yz-$planes through the center of the computational domain are projected to the corresponding rear walls of the box bounding the computational domain. We accomplish this by using the \texttt{Slice} operator in VisIt to get the two-dimensional slices and the \texttt{Transform} operator to put the slices on the walls of the bounding box.

To show the shock surface, we use a \texttt{Contour} plot of the quantity $\mathrm{MachNumber}=2.0$. We set the opacity of this plot to $25$ percent to allow us to see through the shock surface. We also show the isosurface of entropy value $0.15$ using the \texttt{Contour} plot. These two plots display the three-dimensionality of the shock structure from a fixed viewing angle. 

For every output of a single time slice from the simulations, a visualization frame with the above plots is created. Multiple frames are then assembled to make a visualization animation (e.g. movie) from a single simulation. To show progress through the animation, and to give a better sense of the volume of the shocked material, we also plot shock volume averages using 1D \texttt{Curve} plots, with solid lines to indicate the shock trajectory thus far in the simulation and dotted lines to indicate the future trajectory.
%FIXME:(the data for this plot were generated using post-processing analysis as discussed in $\S$ \ref{subsec:simulations}).
The two frames (entropy/fluid plots and shock volume Curve plot) for each time slice are then merged using the image processing tool ImageMagick.\footnote{http://www.imagemagick.org/}

\begin{figure}
\centering
\includegraphics[width=0.5\textwidth]{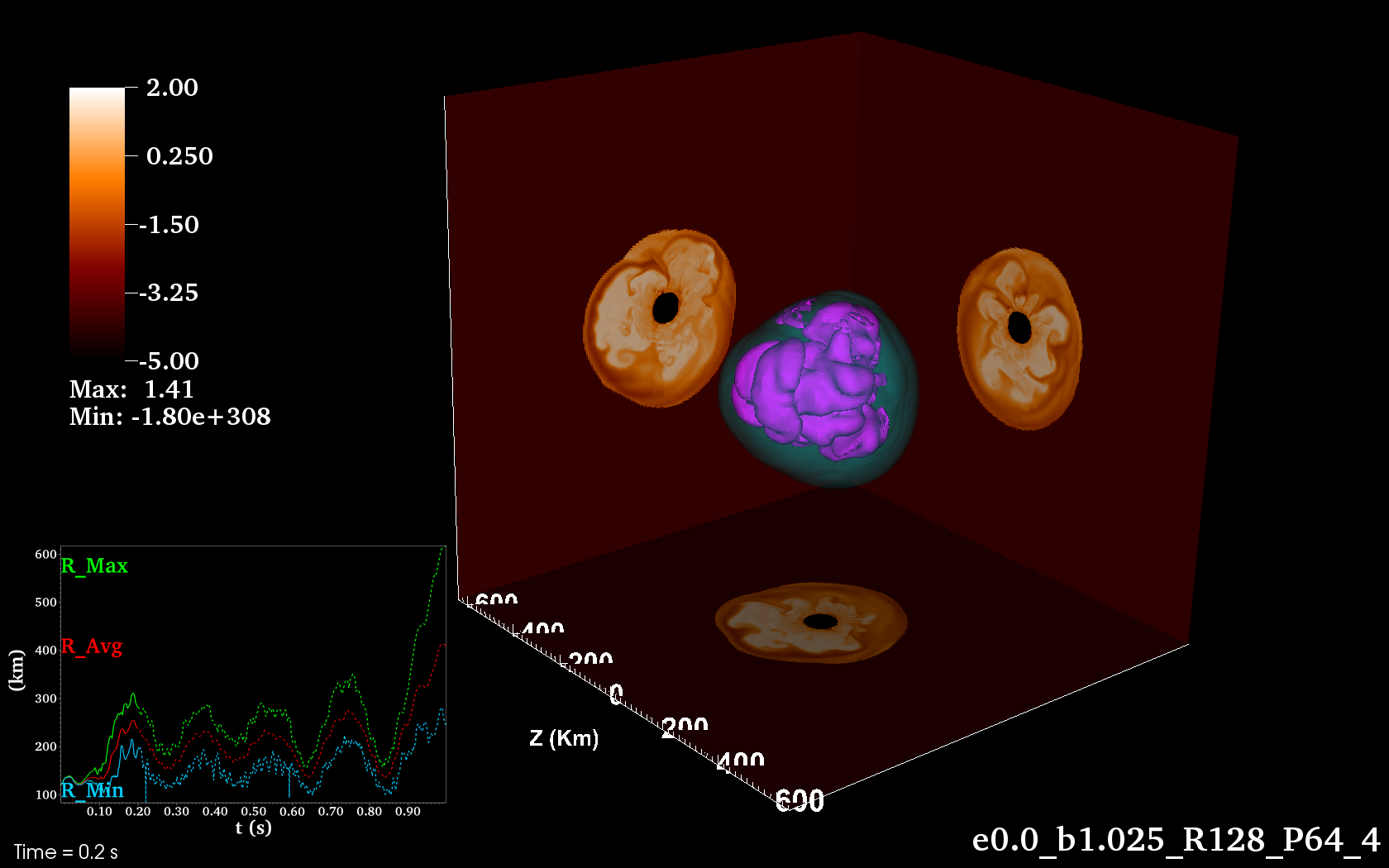}
\includegraphics[width=0.5\textwidth]{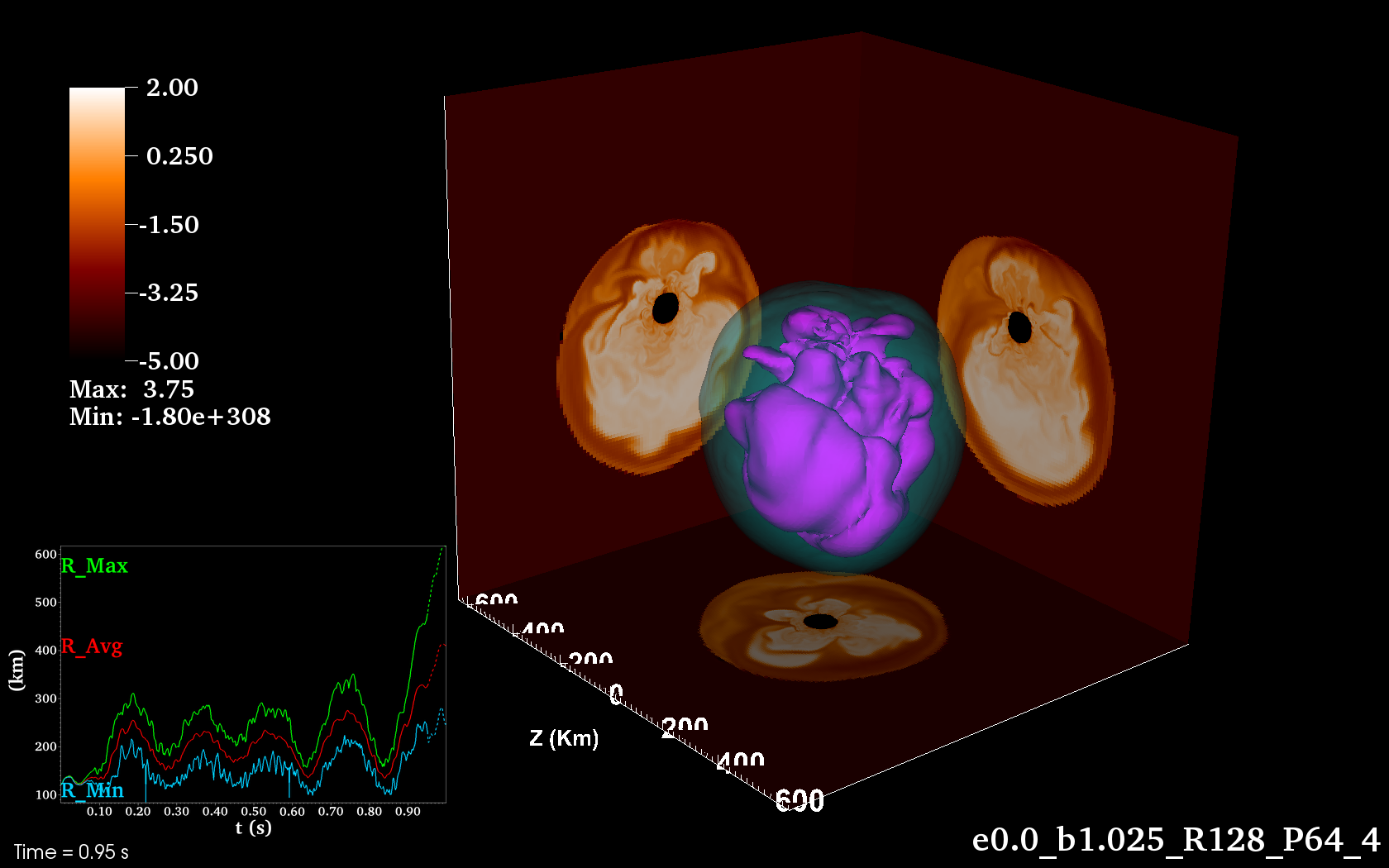}
\caption{SASI-dominated simulation for $B=1.025$ at early time $t=0.2$ s and late time $t=0.95$ s preceding explosion.}
\label{fig:sasi_dominated_vis}
\end{figure}

\begin{figure}
\centering
\includegraphics[width=0.5\textwidth]{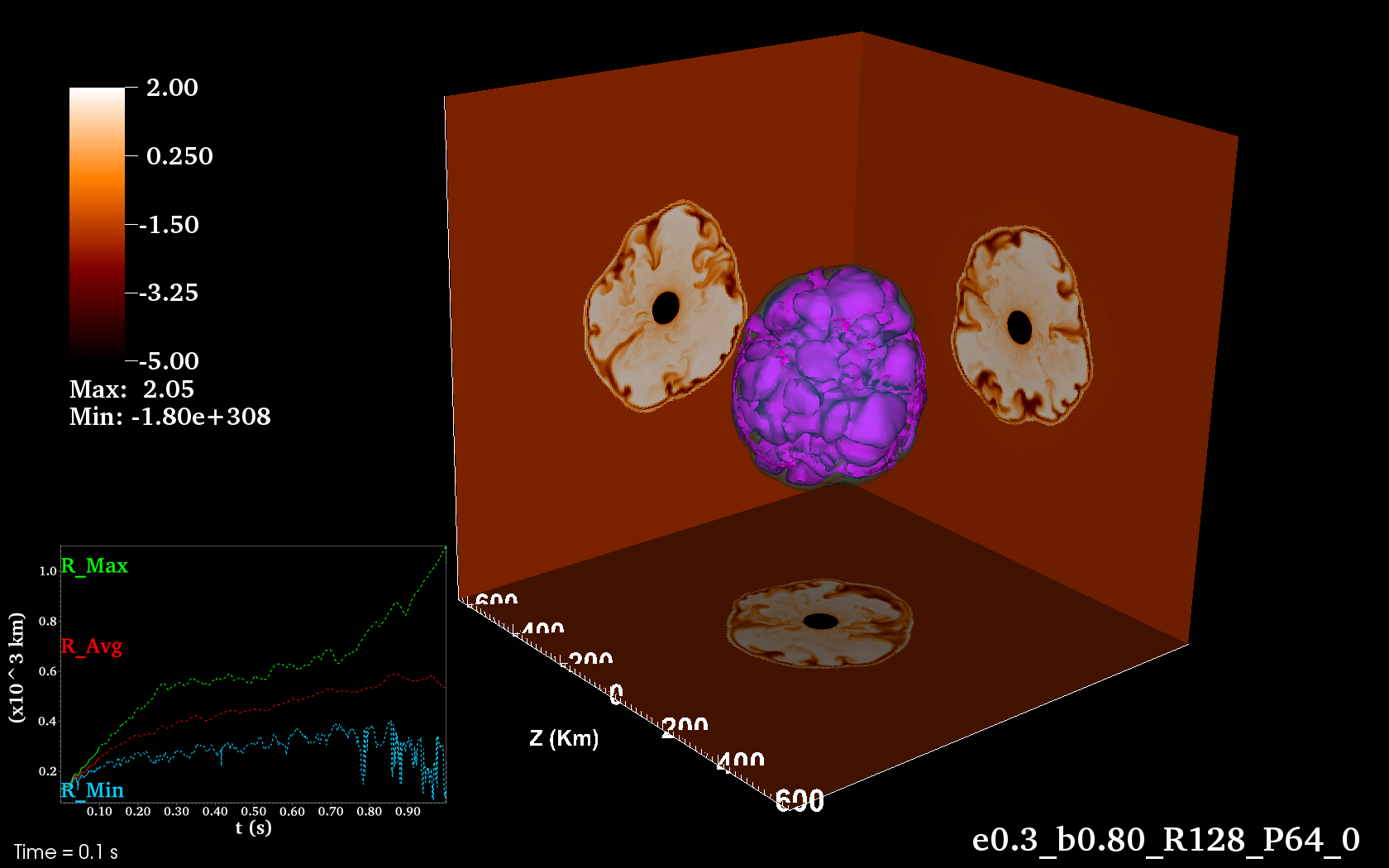}
\includegraphics[width=0.5\textwidth]{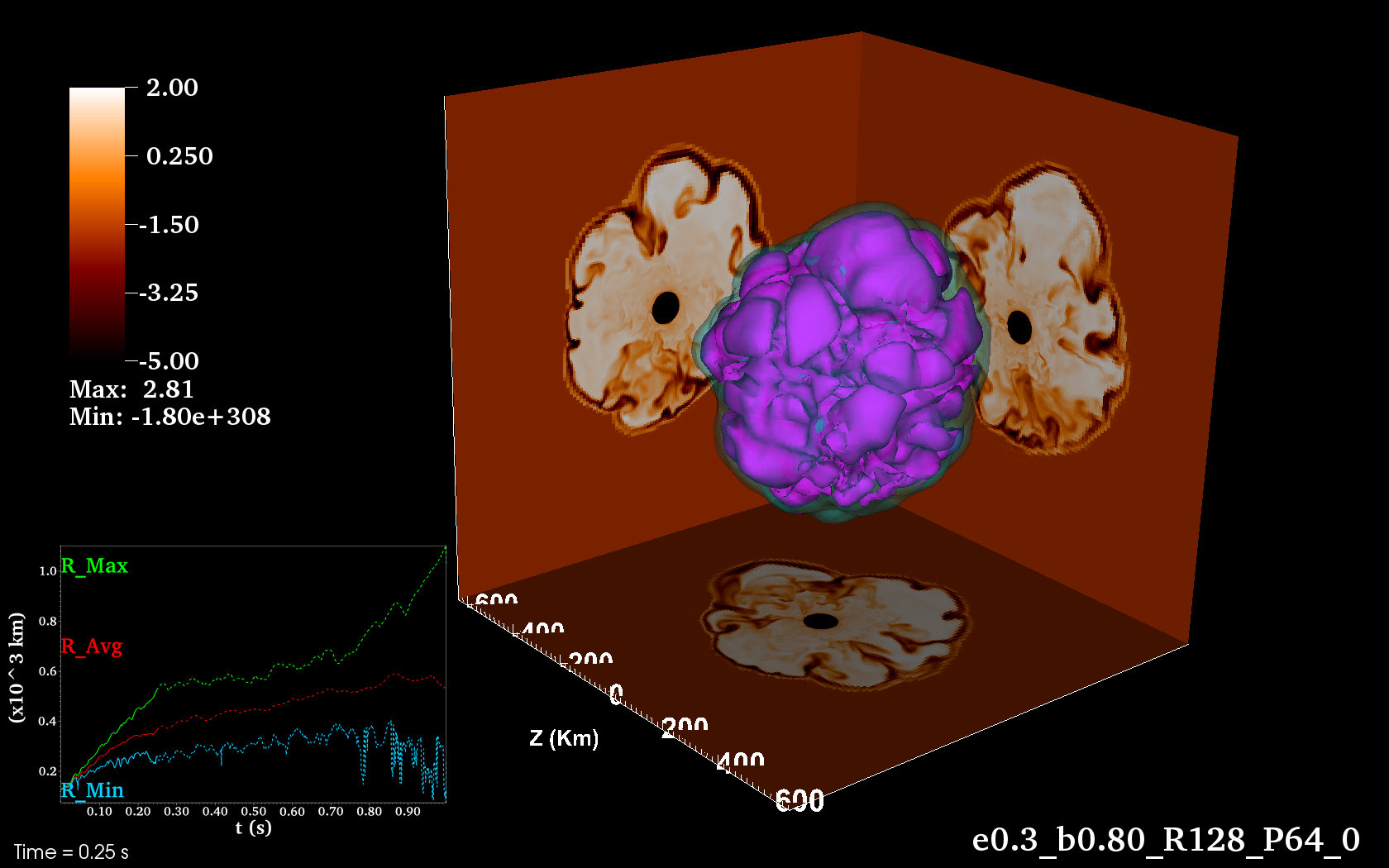}
\caption{Convection-dominated simulation for $B=0.80$ at early time $t=0.1$ s and late time $t=0.25$ s preceding explosion.}
\label{fig:convection_dominated_vis}
\end{figure}

We used supercomputing resources Darter and Beacon at NICS to create the visualization movies from the simulations. Visualization scripts written in Python were used to drive VisIt and generate the plots described above. The VisIt scripts were then run in batch mode on Darter to do parallel rendering and generate PNG frames. Beacon was used to do image processing of the frames and to assemble them into movies using Mplayer software\footnote{https://www.mplayerhq.hu/}. Beacon's large memory capacity and on-node solid-state drives were useful to process the images efficiently. NICS' system-wide Medusa filesystem was very useful in allowing us to use multiple resources for simulations, post-processing, and visualizations without the need to move large amounts of data.

Figures \ref{fig:sasi_dominated_vis} and \ref{fig:convection_dominated_vis} show the visualization described here for one of the simulations at heating value $B=1.025$ for the SASI-dominated model and $B=0.80$ for the convection dominated model. Some visualization movies from the simulations will be presented during the conference.

\section{Conclusion}
\label{sec:conclusion}
In this paper we have presented an abridged description of our simulation results from over a hundred simulations of both convection- and SASI-dominated dynamics in the core-collapse supernova environment with \genasis. A particularly important, previously undiscovered, result is the stochastic behavior of the SASI-dominated models. Because three-dimensional supernova simulations are typically very computationally expensive, it is unheard of to run tens, let alone hundreds of three-dimensional simulations. We were able to do so due to the simplified physics in our setup and the use of multilevel mesh refinement. Yet these models help us to tease out important ingredients of successful supernova explosions. They may also help us interpret results from more sophisticated physical models. Much more detailed analysis from our simulation results is currently still ongoing, and will be the subject of future studies. 

\section*{Acknowledgment}
This material is based upon work performed using computational resources 
Darter \cite{Fahey2014} supported by the University of Tennessee and Oak Ridge National 
Laboratory's Joint Institute for Computational Sciences 
(http://www.jics.utk.edu). Any opinions, findings, and conclusions or 
recommendations expressed in this material are those of the author(s) and do 
not necessarily reflect the views of the University of Tennessee, Oak Ridge 
National Laboratory, or the Joint Institute for Computational Sciences.

%\begin{thebibliography}{9}

%\bibliographystyle{IEEEtran}
\bibliographystyle{unsrt}
\bibliography{references}
%\end{thebibliography}

\end{document}